\documentclass[useAMS,usenatbib]{mn2e}

\usepackage{graphicx}

\title[An Automated System to Classify Stellar Spectra I.]{An Automated System to Classify Stellar Spectra I.}
\author[C. Allende Prieto]{Carlos Allende Prieto\thanks{E-mail:
callende@astro.as.utexas.edu} \\
Department of Astronomy, University of Texas, Austin, TX 78712-1083, USA}

\begin{document}

\date{}

\pagerange{\pageref{firstpage}--\pageref{lastpage}} \pubyear{2002}

\maketitle

\label{firstpage}

\begin{abstract}

Analyses of stellar spectra often begin with the determination 
of a number of  parameters that define a model atmosphere.
%
This work presents a prototype  for an automated spectral classification
system that uses a 150 \AA-wide region around H${\beta}$, 
and applies to stars of spectral types A to K with 
normal (scaled solar) chemical composition. 
The new tool exploits synthetic spectra
 based on plane-parallel flux-constant model atmospheres. The
  input data are high signal-to-noise spectra with
a resolution greater than about 1 \AA.
 The output parameters are forced to agree with an external scale of effective
 temperatures, based on the Infrared Flux Method.
The system is fast -- a spectrum
is classified in a few seconds-- and well-suited for implementation on a web
server. We estimate upper limits to the $1\sigma$ random error
in the retrieved 
effective temperatures, surface gravities, and metallicities as  100 K, 0.3 dex,
and 0.1 dex, respectively.

\end{abstract}


\section{Introduction}

Mass, radius, and luminosity  are some of the most interesting
properties of a star. Unfortunately, it is non-linear combinations of them
that produce quasi-linear changes on a stellar spectrum.
Stellar fluxes are
commonly interpreted in terms of atmospheric temperature, pressure, and 
chemical composition.
In the context of classical flux-constant model atmospheres, these fields
are simply specified with three scalars: the effective temperature, the surface 
gravity, and a solar-scaled metallicity. Nevertheless, extracting the 
three from an observed spectrum is rarely trivial.  
 
A  spectroscopic classification system has been developed independently
of the physical parameters. The MK
system (Morgan, Keenan \& Kellman 1943)
lays out a series of rules to assign spectral classes from medium-low
resolution spectra. This method has the advantage of providing a 
standard reference independent of models. 
However, it is somewhat artificial, in the sense
that the defined spectral classes are obviously  correlated with the  
relevant atmospheric parameters. Moreover, the classical MK system does not
provide for  metal-poor stars. One may note,
as an example, that the metal-poor giant HD 122563, which has an effective 
temperature around 4600 K, has been often 
classified as a late-F or early-G type star.
On the other hand, classification methods based on  physical
parameters are more natural, but model-dependent. 

Both the MK and the {\it physical} classification systems have their own
advantages,  and this may be the reason why they still coexist.
But, the fact  that it is the physical parameters  what is ultimately 
demanded for astronomical applications is shifting most of the recent research  
toward the direct extraction of those quantities.
As recognized by many specialists, repeatability and high speed in
spectroscopic stellar classification can only be achieved by using automatic 
 methods. A recent discussion  of the most used methods can be found in
 Bailer-Jones (2001).

Derived stellar parameters  may differ when determined from  different wavelength
ranges or spectral features. Among other reasons, 
this may be caused by using models that are too
simplistic. As the parameters will be derived from their expected 
effect on the spectra,
  inaccurate predictions, or neglect of 
other relevant parameters, will bias the results. Details in the 
implementation of a classification procedure are also a reason to worry. 
The wide range of effective temperatures that have been assigned to the 
metal-poor subgiant HD 140283 in the recent literature serves as testimony
of the respectable uncertainties still 
involved in the scale of effective temperatures
 (see, e.g., the discussion in Snider et al. 2001). 

Discrepancies produced by interpreting stellar spectra with model atmospheres
that are too simple will decrease as progress in theory takes place. 
 For main-sequence stars, remarkable advances are 
happening as  efforts focus on relaxing the assumptions of 
Local Thermodynamic Equilibrium (LTE) and hydrostatic equilibrium. 
Systematic differences that
arise in the implementation of different methods for spectroscopic 
classification can be controlled by establishing standards. A reference 
implementation would be far more powerful than a  set of standard stars.
Modern information technologies have paved the way for implementing 
a public automatic classification system accessible over the Internet. 
Such open system, if reliable and fast,  could serve as a standard reference. 

\begin{figure*}
\centering
\includegraphics[width=8cm,angle=90]{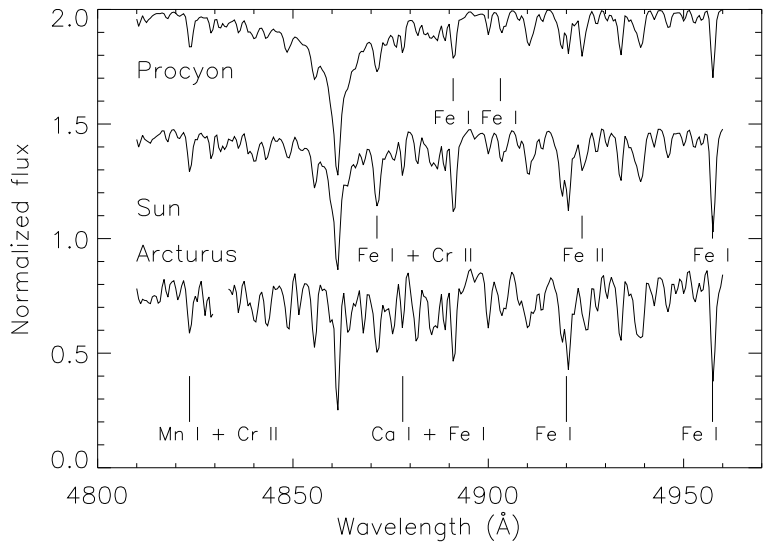}  
\protect\caption[ ]{
Observed fluxes for Procyon, the Sun and Arcturus in the selected window. 
The original very high dispersion observations have been convolved with a 
Gaussian profile with a FWHM of 1.0 \AA. The observations in the McDonald 
atlas of Procyon (Allende Prieto et al. 2002) have been continuum normalized
by P. S. Barklem (see Barklem et al. 2002 for more details). A missing interval
in the spectrum of Arcturus is severely affected by telluric lines.
\label{f1}}
\end{figure*} 

This paper discusses the first steps toward the implementation of a 
prototype for an open classification system. This work will only deal with
a section of the HR diagram; stars with spectral types A to K and 
 scaled solar metal abundances. 
Section 2 describes the selected
wavelength  band,  \S 3 the working parameters space, and \S4 
provides details of the implementation. Section 5 is devoted
to connecting the parameters derived from spectroscopy 
to a widely accepted scale
of effective temperatures based on the Infrared Flux Method, 
and checking the performance of the method.
Section 6 concludes with a summary of present results and ideas for 
subsequent work.

\section{Wavelength range}

When using equivalent widths and classical flux-constant model atmospheres
in abundance analyses,
the relevant parameters can be usually 
reduced to four: the stellar effective
temperature ($T_{\rm eff}$), the surface gravity  ($g$), the micro-turbulence
($\xi$), and the metal abundances -- commonly considered proportional to the
iron abundance. 
For stars with spectral types A-K, 
the most useful and accessible observational probes for these 
parameters are the flux distribution and the spectral lines:
 excitation and ionization equilibrium of metals, 
the pressure-enhanced wings of strong metal lines, 
and the hydrogen lines. Different procedures to derive 
stellar parameters rely on one or several of these indicators.

At this point we set aside potential  
difficulties to model some of the spectral features mentioned above. 
We refer the reader to the
papers by 
Dragon \& Mutschlecner (1980),  Castelli,  Gratton \& Kurucz (1997), or 
Bell, Balachandran \& Bautista (2001)
regarding the modelling of the continuum flux; 
Th\'evenin \& Idiart (1999) or Asplund et al. (2000) 
on the calculation of Fe I line profiles; and
Fuhrmann, Axer \& Gehren (1996),  Gardiner,
Kupka \& Smalley (1999), Barklem, Piskunov \& O'Mara (2000), or 
Cowley \& Castelli (2002) on modelling Balmer lines.
With the ultimate goal of deriving chemical abundances, most spectroscopic
observations aim at providing reliable line profiles or equivalent widths for
 lines of weak-to-moderate strength. Ideally, one would use the same type of 
spectra to determine
the stellar atmospheric parameters. 
In addition, accurate spectrophotometry is challenging
and  highly vulnerable to reddening.
Forcing the excitation and
ionization equilibrium balance for metal lines is, in most cases,
 insufficient to
reliably determine the quartet ($T_{\rm eff}$, $\log g$,  
[Fe/H]\footnote{[Fe/H] $= \log \frac{N({\rm Fe})}{N({\rm H})} - \log 
\left[\frac{N({\rm Fe})}{N({\rm H})}\right]_{\odot}$, where $N({\rm E})$
represents the number density of the element E.}, $\xi$). 
Therefore, we resort to a second feature, the  Balmer lines.

We selected a continuous spectral window around H${\beta}$: 4810--4960 \AA. 
This wavelength range represents a balance in many aspects. 
It is red enough that the
continuum opacity is well described by H and H$^{-}$ for the stars under
consideration, avoiding the difficulties of dealing with much more complicated 
(metal) opacities, and it is blue enough that the presence of spectral lines 
makes possible a reliable determination of the metal abundance -- even in 
metal-poor stars.

\section{Working domain}

Use of equivalent widths, quantifying the strength of 
a spectral line by a single number, represents a loss of information.
Use of line profiles introduces more variables in the analysis through
 the different line broadening mechanisms. 
Accounting for the broadening involves a number
of difficulties in the practical implementation of a classification algorithm,
although it also comes with extra information, e.g. projected rotational 
velocity, or instrument spectral resolution. At this point we wish to
restrict the classification to ($T_{\rm eff}$, $\log g$,  [Fe/H], $\xi$) and,
therefore, we use a fixed resolving power $R \equiv \lambda/\delta\lambda
\simeq 5000$. We covered the selected spectral range with 301 points, 
equally spaced in wavelength (every 0.5 \AA).
 Our choice of resolution is a compromise: low enough to make rotational
and macro-turbulent broadening in late-type stars negligible, and 
high enough to be able to recover information on the stellar 
atmospheric parameters. Fig. 1 shows observations for the 
Sun (solid; Kurucz et al.  1984), Procyon
(dashed; Allende Prieto et al. 2002), and
Arcturus (dash-dotted; Hinkle et al. 2000). The most relevant features
have been identified in the figure.

The range for each of the atmospheric parameters was selected to avoid some
extreme conditions where  classical model atmospheres in general, or those
used here in particular, are expected, or known, 
to fail: cool temperatures at which the contribution of molecules to the
equation of state is incomplete in the models; 
hot temperatures at which departures from LTE
are important for the atmospheric structure; or too extended an atmosphere 
that the  plane-parallel approximation is inadequate. The selected domain is:

\begin{eqnarray}
\begin{tabular}{cccccc}
4500 & $\le$ & $T_{\rm eff}$ & $\le$ & 8000 & K\\
2.0 & $\le$ & $ \log g$ & $\le$ & 5.0 & dex \\
$-4.5$ & $\le$ & [Fe/H] & $\le$ &  $+0.5$ & dex \\
0 & $\le$ &  $\xi$ & $\le$ &  2 & km s$^{-1}$. \\
\end{tabular}
\end{eqnarray}

\noindent Arcturus is probably cooler than our lower limit for $T_{\rm eff}$
[Griffin \& Lynas-Gray (1999) assign $4290 \pm 30$ K to this star], but
it serves the purpose of showing 
an example of the coolest spectra in our working sample. 
As  will become clear later, our raw $T_{\rm eff}$s are
systematically higher than other scales and thus a star with this 
effective temperature
is technically within the limits of our grid.

Extensive testing showed that the flux in the selected spectral window 
satisfies a one-to-one relationship with most of the parameters space. 
In other words, for a given combination of the four atmospheric parameters
considered, the resulting flux in this window is unique. Approaching the
extreme metal-poor limit, below [Fe/H] $< \sim -3$, degeneracy is unavoidable,
as metal lines vanished, and so does the information on metallicity, 
gravity and microturbulence. The  flux in the selected
window is not equally sensitive to changes in the different stellar parameters.
Changes in $T_{\rm eff}$ affect the most the spectrum, followed by variations
in the metal abundance. The effect of changes in the surface gravity 
is mainly felt  through the different sensitivity  of lines of
neutral and ionized metals to pressure, and it is more subtle than 
the response to $T_{\rm eff}$ or [Fe/H]. 

\section{Implementation}

To find the  set of parameters that best reproduces an observed spectrum,
we need to choose an algorithm. We want to minimize (or 
maximize) a function of the stellar parameters. This will require us to 
evaluate a function -- thus compute synthetic spectra -- a number of
times. This number can be very large, and therefore a strategy 
to reduce computing time is needed. We have tackled this problem
by computing a discrete grid and
interpolating. 
We adopted the following increments for the four parameters involved: 
500 K in $T_{\rm eff}$, 1.0 dex in $\log g$, 1.0 dex in [Fe/H], 
and 1.0  km s$^{-1}$ in $\xi$. These were
chosen to keep the changes in the spectrum small enough so that a fast 
multilinear interpolation would provide a reasonable approximation. 
Interpolation errors can reach up to 2 \% for the warmer stars in the grid,
but up to 7 \% for the coolest metal-rich stars. These errors are smaller than
the precision with which the real spectra can be reproduced, and it was 
later verified that use of a finer grid does not improve the performance of
the classification method.
We used a 
 genetic algorithm (GA) to solve our minimization problem. GAs are suitable 
 for solving global optimization problems in complex landscapes where
 local extrema can confuse simpler algorithms 
 (see Charbonneau 2002 for an informal introduction). 

The grid of synthetic spectra was based on non-overshooting
 Kurucz (1993) model atmospheres. These models include
mixing-length convection with $\alpha=1.25$, and $\xi=2.0$ km s$^{-1}$. The
radiative transfer equation was solved with the code 
{\tt Synspec} (Hubeny \& Lanz 2000), using 
very simple continuous opacities: H, H$^{-}$, Rayleigh 
and electron scattering (as described in Hubeny 1988). 
An atomic line list was prepared with the data obtained
from the Vienna Atomic Line Database (VALD; Kupka et al. 1999). This line list
includes 7169 lines that are expected to contribute to the opacity 
in a solar-like atmosphere.
The computed spectra were degraded to a resolving
power of about 5000, sampled with a common
wavelength vector, and normalized. 

\begin{figure*}
\centering
\includegraphics[width=6cm,angle=90]{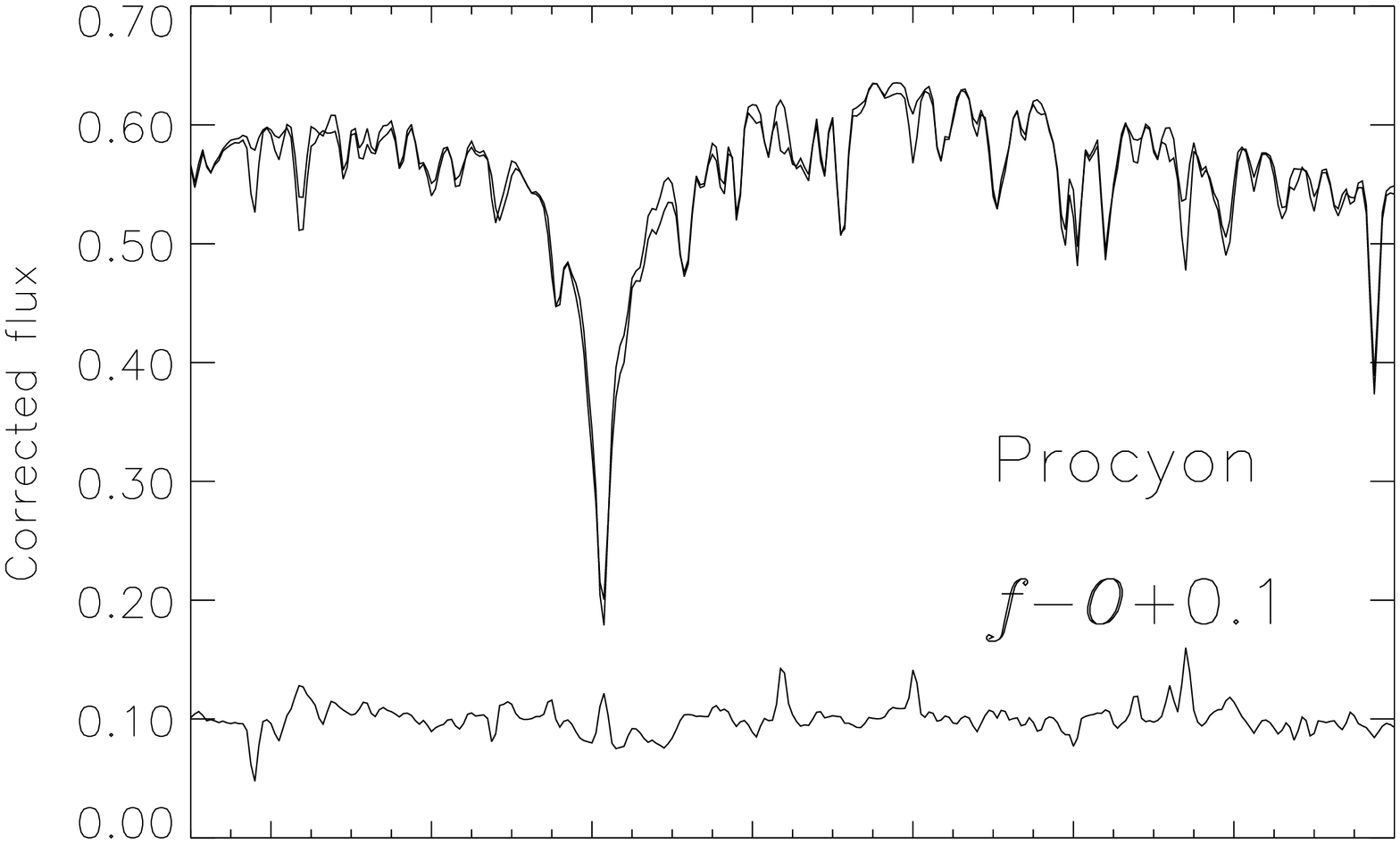}  
\includegraphics[width=6cm,angle=90]{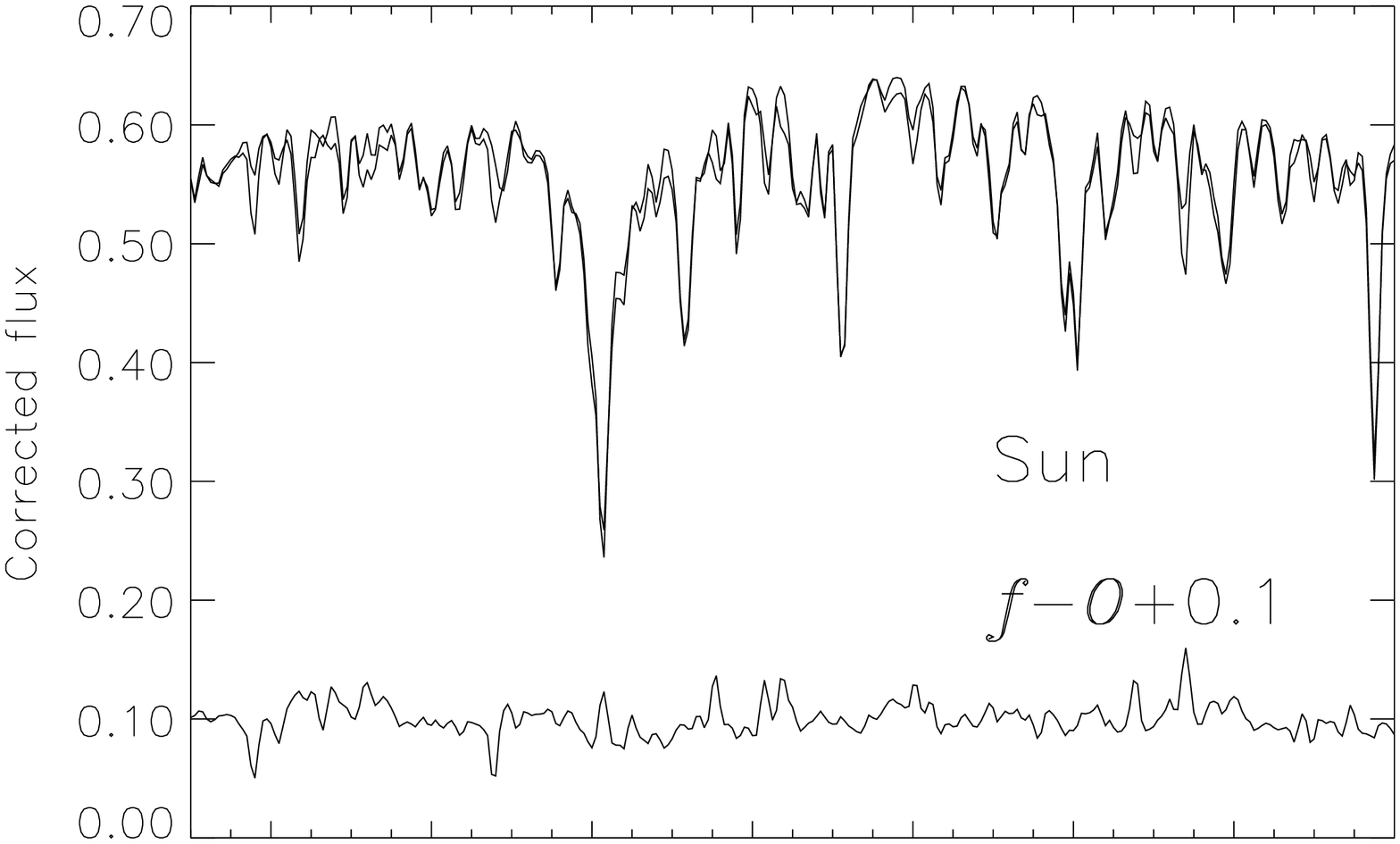}  
\includegraphics[width=6cm,angle=90]{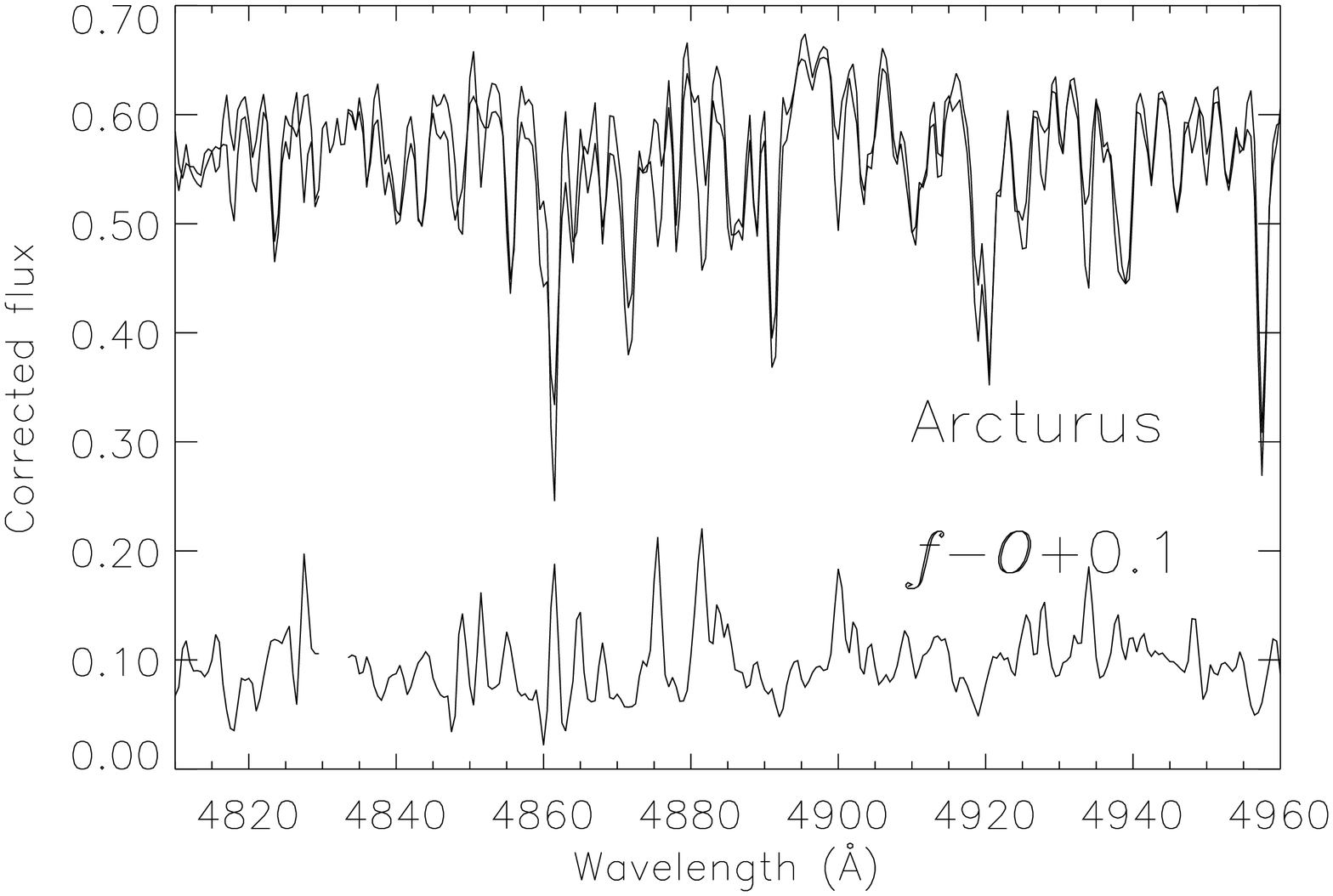}  
\includegraphics[width=6cm,angle=90]{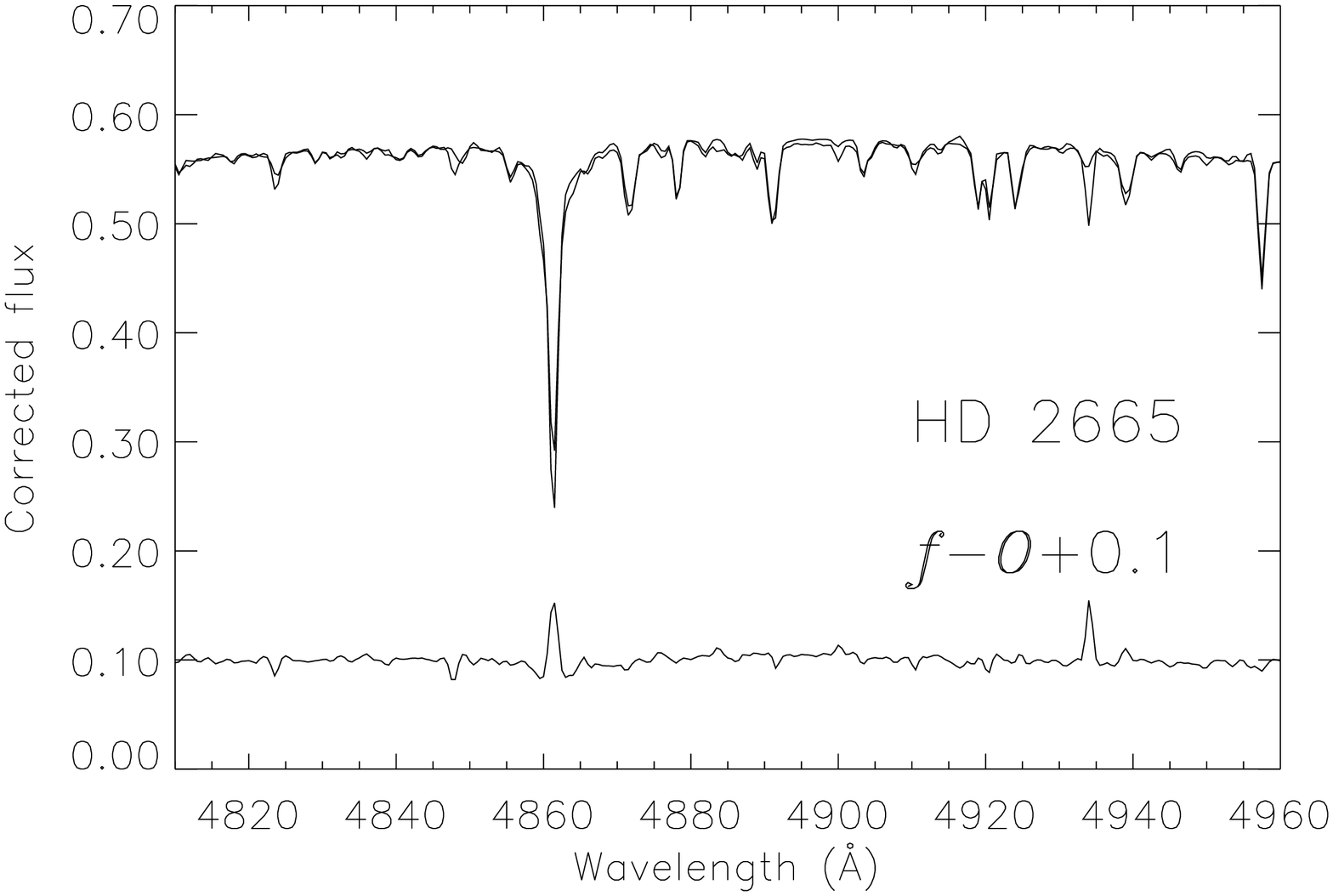}
\protect\caption[ ]{
Comparison between the observed ($O$) and  (linearly interpolated) synthetic 
($f$) corrected fluxes for the Procyon (F5 IV), the Sun (G2 V),
 Arcturus (K1.5 III), and the metal-poor HD 2665 (G5 III; [Fe/H] $\simeq -1.9$). 
 The difference of the two vectors is also plotted.
\label{f2}}
\end{figure*}

The presence of a very broad line in the spectral window ($H{\beta}$)
makes normalization difficult. We have adopted a straightforward unsupervised
polynomial normalization. Although the results of this scheme would visually
displease most stellar spectroscopists (see Fig. \ref{f2}), 
this simple procedure can be carried 
out quickly, and repeatability is easily achieved for a given set 
($T_{\rm eff}$, $\log g$, [Fe/H], $\xi$). In addition, the fluxes were also 
divided by a constant, 1.8, to enforce flux values between 0 and 1
 -- a requirement of the GA software.
 
We implemented a FORTRAN routine to perform multilinear interpolation in the
four parameters under consideration. This routine is the interface between
the grid of synthetic spectra and the genetic algorithm. The function we chose 
to maximize is:

\begin{equation}
1 - \sum_i w_i (f_i-O_i)^2
\end{equation}

\noindent where {\it  f} is the vector 
of interpolated synthetic flux, {\it O} is the vector of 
 observations, and the index {\it i} indicates a particular wavelength bin. We
 adopted

\begin{equation}
w_i = (f_i^{\odot}-O_i^{\odot})^{-2}/10^4,
\end{equation}

\noindent as derived from the solar spectrum and a synthetic flux 
calculated with solar parameters,  but reset to

\begin{eqnarray}
\begin{tabular}{cccccccc}
 $\biggl\{$  & 
\begin{tabular}{c}
 	   $10$     \\
 	   $0$      \\
\end{tabular} & 
\begin{tabular}{c}
 	    if $w_i > 10$\\   
 	    if $1.0 > w_i$. \\
\end{tabular} 
& & & & 
\\
\end{tabular}
\end{eqnarray}

The many necessary multilinear interpolations are very fast, as the grid
of previously computed synthetic spectra is kept in memory. In
fact, a non-negligible fraction of the time is invested in loading those data. 
We adopted a publicly available GA software\footnote{Available from 
{\tt http://cuaerospace.com/carroll/ga.html}},
due to  D. L. Carroll (see, e.g., Carroll 1996). The default parameters were 
kept, namely, a micro-GA  with uniform crossover. The GA  
 was run for 500 generations. This number was chosen from inspection of the
 convergency curve for a limited number of stars, but it was later verified
 that increasing this value to 2000 generations did not produce any 
 improvement in the performance. Classification of a single
spectrum takes about 3 seconds on a Sun Ultra5.  Fig. 2 illustrates the 
agreement between the observed and the matched synthetic
spectra for Procyon, the Sun,  Arcturus, 
and the metal-poor star HD 2665 ([Fe/H] $\simeq -1.9$). The spectrum of
HD 2665 was obtained from the {\it Elodie} library (Prugniel \& Soubiran 2001).

\begin{figure*}
\centering
\includegraphics[width=7cm,angle=90]{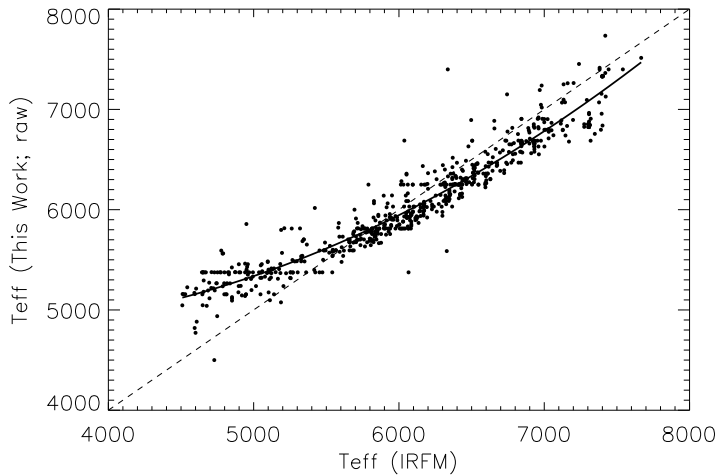}  
\protect\caption[ ]{
The raw effective temperatures derived from fitting the spectral region
4810--4960 \AA\ compared to those derived from the $(B-V)$ calibrations of
Alonso et al. (1996, 1999). The solid line is a second-order least-squares
polynomial fit that we use to correct our $T_{\rm eff}$ scale.
\label{f3}}
\end{figure*} 

\begin{figure*}
\centering
\includegraphics[width=10cm,angle=0]{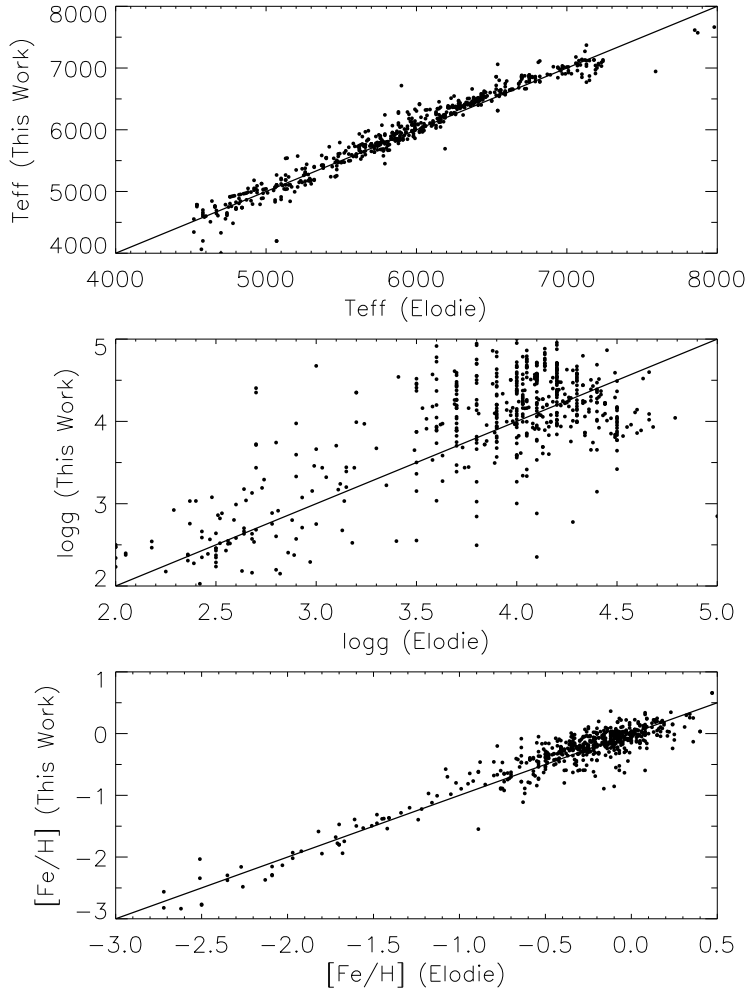}  
\protect\caption[ ]{
Comparison between the stellar parameters selected (mainly from the literature)
for the Elodie stars (Prugniel \& Soubiran 1999 and references therein) with
those from spectral fitting derived in this work.
\label{f4}}
\end{figure*} 

\begin{figure*}
\centering
\includegraphics[width=7cm,angle=90]{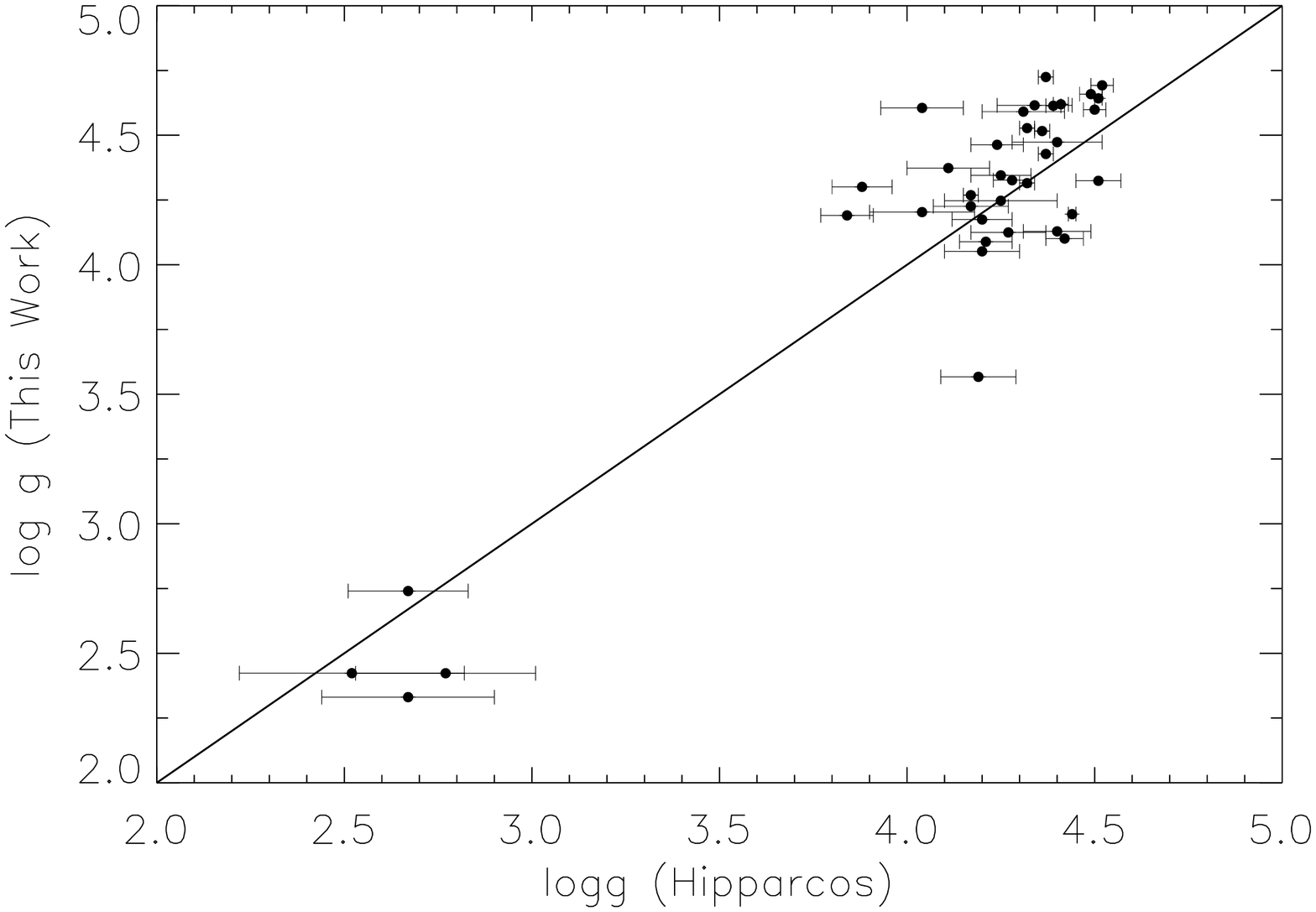}  
\protect\caption[ ]{
Comparison between the gravities determined from Hipparcos' parallaxes by
Allende Prieto \& Lambert (1999) with
those from spectral fitting derived in this work. 
\label{f5}}
\end{figure*}

\section{Connecting our stellar parameters to the IRFM $T_{\rm eff}$ scale}

Detailed modelling of H${\beta}$ is an issue. Even though the wings of 
Balmer lines are considered to form very close to LTE conditions, that
does not apply to their cores. A tougher
complication is the fact that Balmer lines are commonly affected by the
temperature distribution in the deepest atmospheric layers, which  
for late-type stars are significantly influenced by convection. Convection is
typically treated using a mixing-length formalism which implies a choice
for one or several parameters. This represents an important additional
source of uncertainty in the derived parameters -- very likely a systematic
bias in the derived $T_{\rm eff}$. Such a bias in $T_{\rm eff}$,
 because of the tight coupling between
$T_{\rm eff}$, $g$, and [Fe/H], will also produce a 
 bias in the other two parameters.
In our view, the best possible option is to anchor our spectroscopic 
$T_{\rm eff}$ scale to a more reliable scale. The systematic effect in
$g$ and [Fe/H] implied by the necessary correction to our derived 
$T_{\rm eff}$ can be easily predicted. Empirical studies have shown that
a shift in an adopted $T_{\rm eff}$ will, through the iron 
excitation-ionization
balance, translate to a shift in $\log g$:

\begin{equation}
\Delta \log g \simeq \Delta T_{\rm eff}/466.
\end{equation}

\noindent (Allende Prieto 1998), and a correction to [Fe/H]:

\begin{equation}
\Delta {\rm [Fe/H]} \simeq \frac{\Delta \log g}{3.0}
\end{equation}

\noindent (e.g. Gray 1992; Allende Prieto et al. 1999).

As we will be applying this correction to  our $T_{\rm eff}$ scale,
and the sensitivity to $T_{\rm eff}$ of the selected spectral region 
relies largely on the $H{\beta}$ profile, there is no need to make 
 a serious emphasis on the accuracy of the calculated absorption profile for
H${\beta}$. We can make use of some approximations, and take advantage of
a reduced computing time. In particular, we  use
an approximate broadening treatment described in the Appendix B
 of Hubeny, Hummer \& Lanz (1994). 
As this approach  underestimates the
width of the solar H${\beta}$ profile, we apply a zeroth order correction, 
using twice the default value. This modification is not strictly 
necessary, as the 
correction to be applied later to the $T_{\rm eff}$ scale would be able to fix this zero point
as well.

We decided to anchor our spectroscopic $T_{\rm eff}$ scale to that of
Alonso, Arribas \& Mart\'{\i}nez-Roger (1996, 1999), which is based 
on  the Infrared Flux Method, as modelled with fluxes from Kurucz (1993)
 model atmospheres.
A  library of high-resolution spectra recently published by 
Prugniel \& Soubiran (2001) is used  as testing field for 
our classification scheme. This library, hereafter the {\it Elodie} library,
consists of more than 908 spectra from 709 stars spanning a large fraction of, 
and in some instances exceeding,  our parameter space. We determine the
$T_{\rm eff}$ for each star using the Alonso et al. calibrations for $(B-V)$.
These calibrations use $(B-V)$ and [Fe/H], and both parameters were adopted from
those given in the {\it Elodie} library. The division of the stars 
in IV-V or I-III
classes, in order to select the appropriate  IRFM calibration, 
is based on the
gravities provided in the {\it Elodie} library, setting the division line
 at $\log g = 3.8$.

The spectra in the library with a resolution of $R=10,000$ are convolved 
with a Gaussian profile with a FWHM of 1.0 \AA, and then 
fed to our classification system  to find a
best-fitting vector ($T_{\rm eff}$, $g$, [Fe/H], $\xi$). Then,
the values derived for 
$T_{\rm eff}^{raw}$ are compared against those from the IRFM 
calibrations. The two scales are confronted in Fig. 3. A second-order
least-squares polynomial fit is adopted to anchor the derived 
$T_{\rm eff}^{raw}$ to the IRFM scale:

\begin{equation}
T_{\rm eff}^{raw} =  5783  - 0.6686~ T_{\rm eff}^{IRFM} +
0.0001159 ~(T_{\rm eff}^{IRFM})^2,
\end{equation}

and this correction is also translated to $\log g$, and [Fe/H], based on the
correlation expected for the excitation-ionization balance (see Eqs. 5 and 6).
Fig. 3 reveals a tendency for some stars to clump at certain values of 
$T_{\rm eff}$, in particular at $\simeq 5375$ K. Too aggressive
 a choice for the steps in the grid parameters (with
the implied errors in the linear interpolation) is not  
to blame for this systematic
effect, which is related to caveats involved in the
implementation of GAs. Noticeably, the scatter is not symmetrically distributed about
the adopted mean relationship, defined by the polynomial fit. This is actually
what we expect when reddening is not negligible, introducing significant
errors in the photometric $T_{\rm eff}$ [which are based on unreddened values
for $(B-V)$]. Addressing these 
and other issues exceeds the scope of this exploratory study.

The corrected $T_{\rm eff}$, $\log g$, and [Fe/H] values 
can be directly compared
to those given in the {\it Elodie} library. The mean difference 
and the  $\sigma_{rms}$ (607 spectra) are

\begin{eqnarray}
\begin{tabular}{cc}
$T_{\rm eff}$: & $37 \pm 150$ K \\
$\log g$:  & $0.16 \pm 0.52$ dex \\
${\rm [Fe/H]}$: & $0.02 \pm 0.18$ dex \\
\end{tabular}
\end{eqnarray}

\noindent and Fig. 4 compares the two sets of parameters.

A contribution to the error bars is connected with the parameters 
adopted for the {\it Elodie} library, mainly compiled from the literature. 
The library also provides estimates of the reliability of
the adopted values. Restricting the comparison to the spectra with the 
most trusted 
parameters\footnote{Using the library's code: q($T_{\rm eff}$)=4, q($\log g$)=1, q([Fe/H])=4},
we find (71 spectra)

\begin{eqnarray}
\begin{tabular}{cc}
$T_{\rm eff}$: & $21 \pm 102$ K \\
$\log g$: & $0.07 \pm 0.37$ dex \\
${\rm [Fe/H]}$: & $0.02 \pm 0.10$ dex. \\
\end{tabular}
\end{eqnarray}

The $\sigma_{rms}$ for the same stars between the IRFM $T_{\rm eff}$s, and
those we derived is 80 K.
The  agreement for $T_{\rm eff}$ and [Fe/H] is satisfying, but it is not so
for $\log g$. As explained in \S 4, the sensitivity of the spectrum to this
parameter is not nearly as high as to the others. 
We find that 36 stars from the last set are included in the
determination of stellar parameters by Allende Prieto \& Lambert (1999)
based on the comparison of  observed colors and parallaxes
 with evolutionary models. The mean and median uncertainties for the
 reference gravities are both 0.08 dex. For these
 stars, we find a more satisfactory mean difference and $\sigma_{rms}$
 in $\log g$: $0.05 \pm 0.28$ dex, as shown in Fig. \ref{f5}, which leads us to 
 believe that 0.3 dex is a  reasonable estimate for our 
 random errors in gravity.

\section{Conclusions and Future applications}

We have implemented a spectroscopic classification algorithm that provides
 estimates for $T_{\rm eff}$, $\log g$, [Fe/H], and $\xi$ based on the 
 observations with a resolving power $R \ge 5000$ in the 
 spectral range 4810--4960 \AA. The classification system is based on
 synthetic spectra calculated with 
 classical flux-constant model atmospheres, and it is anchored to the
 photometric calibrations of effective temperature derived by 
 Alonso and collaborators 
 (Alonso et al. 1996, 1999) based on the
  Infrared Flux Method. By using the Elodie spectroscopic library (Prugniel
  \& Soubiran 2001) and the gravities determined by 
  Allende Prieto \& Lambert (1999)
 for nearby Hipparcos stars, we derive upper limits to the
 uncertainties in the retrieved $T_{\rm eff}$, $\log g$, and [Fe/H], as
 100 K, 0.3 dex, and 0.1 dex, respectively.
 
 Our classification algorithm  can be used on spectral types A to K, 
 for main-sequence and evolved stars with gravities as low as $\log =2$, 
 and all metallicities. The system 
 is able to classify a stellar spectrum in only
 3 seconds on a modern workstation. Work is in progress to improve the
 accuracy of the synthetic spectra by using  a
 more realistic line absorption profile for H${\beta}$. We also plan on
 varying the parameters that affect the performance of the employed GA, and
testing different optimization algorithms. 
An important question to answer is how the performance of our system improves
 or degrades with spectral resolution and signal-to-noise ratio. Different
 spectral ranges should be explored. Spectral bands with lines of
 metals whose abundances do not scale well with iron should be avoided, unless
 more parameters are included in the search, which is certainly feasible. 

Different classification algorithms for stellar spectra 
have been tested in the literature. 
In particular, artificial neural networks hold
the promise for the highest speeds, which may be critical for problems
involving a large number of free parameters (see, e.g. Bailer-Jones 2001; 
Snider et al 2001). When the number of parameters is limited, like in the 
spectroscopic classification considered here, GAs have the advantage
of not requiring training. This, in turn, allows us to explore different 
strategies, such as the selection of the spectral range or the resolving power,
 very quickly, while keeping the search global.
 
 Our final goal is to provide
 a web interface and make this or a similar system publicly available. 
 Future  work will target hotter spectral types, using
 calculations based upon non-LTE model atmospheres. Extension to the bottom
 of the main-sequence and beyond could follow, although theoretical modelling
 of cool atmospheres  has not yet reached the same level of maturity as for
 warmer stars. 
 The adopted strategy can easily accommodate future improvements 
 in model atmospheres and spectral synthesis.

\section{Acknowledgments}

Paul Barklem, Tim Beers, Norbert Christlieb, David Lambert,  
Chris Sneden, and Ted von Hippel are  thanked for inspiring
discussions. I am obliged to David Carroll for 
making his GA software publicly available, and to the {\it Elodie} team for
sharing their database of stellar spectra.
NSF  support (grant AST 00-86321) is gratefully  acknowledged.

\bsp

\label{lastpage}


\begin{thebibliography} {}

\bibitem[]{} Allende Prieto C., 1998, PhD Thesis, Universidad de La Laguna, La 
Laguna, Instituto de Astrof\'{\i}sica de Canarias

\bibitem[Allende Prieto, Asplund, L{\' o}pez, \& 
Lambert(2002)]{2002ApJ...567..544A} Allende Prieto C., Asplund M., 
Garc\'{\i}a L{\'o}pez R.J., Lambert D.L., 2002, ApJ, 567, 544 

\bibitem[Allende Prieto, Garc{\'i}a L{\' o}pez, Lambert, \& 
Gustafsson(1999)]{1999ApJ...527..879A} Allende Prieto C., Garc\'{\i}a 
L{\'o}pez R.J., Lambert D.L., Gustafsson B.,\ 1999, ApJ, 527, 879 

\bibitem[Allende Prieto \& Lambert(1999)]{1999A&A...352..555A} Allende 
Prieto C., Lambert D.L.,\ 1999, A\&A, 352, 555 

\bibitem[Alonso, Arribas, \& Martinez-Roger(1996)]{1996A&A...313..873A} 
Alonso A., Arribas S.,  Mart\'{\i}nez-Roger C.,\ 1996, A\&A, 313, 873 

\bibitem[Alonso, Arribas, \& Mart{\' 
i}nez-Roger(1999)]{1999A&AS..140..261A} Alonso A., Arribas S.,  
Mart\'{\i}nez-Roger C.,\ 1999, A\&AS, 140, 261 


\bibitem[Asplund et al.(2000)]{2000A&A...359..729A} Asplund M., Nordlund 
{\AA}., Trampedach R., Allende Prieto C., Stein R.F.,\ 2000, A\&A, 
359, 729 


\bibitem [] {} Bailer-Jones C.A.L., 2001, in  
	Gupta R.,  Singh H. P.,  Bailer-Jones C.A.L., eds, Proc.
	Automated data analysis in Astronomy. Narosa Publishing
	House, New Delhi, India, p. 83

\bibitem[Barklem, Piskunov, \& O'Mara(2000)]{2000A&A...363.1091B} Barklem 
P.S., Piskunov N.,  O'Mara B.J.,\ 2000, A\&A, 363, 1091 

	
\bibitem[Barklem et al.(2002)]{2002A&A...385..951B} Barklem P.S., 
Stempels H.C., Allende Prieto C., Kochukhov O.P., Piskunov N.,  
O'Mara B.J.,\ 2002, A\&A, 385, 951 

\bibitem[Bell, Balachandran, \& Bautista(2001)]{2001ApJ...546L..65B} Bell 
R.A., Balachandran S.C.,  Bautista M., \ 2001, ApJL, 546, L65 

\bibitem[]{} Carroll, D.L., 
1996,, in 
 Wilson H.B.,  Batra R.C.,  Bert C.W.,  Davis A.M.J.,  Schapery R.A.,  
Stewart D.S., Swinson F.F., eds, Genetic algorithms and optimizing 
chemical oxygen-iodine 
lasers, Developments in Theoretical and Applied Mechanics, Vol. XVIII, 
School of Engineering, The University of Alabama,  p. 411

\bibitem[Castelli, Gratton, \& Kurucz(1997)]{1997A&A...318..841C} Castelli 
F., Gratton R.G., Kurucz R.L.,\ 1997, A\&A, 318, 841 

\bibitem[]{} Charbonneau P., 2002, NCAR Technical Note 450+IA, High Altitude 
Observatory, NCAR, Boulder, Colorado

\bibitem[Cowley \& Castelli(2002)]{2002A&A...387..595C} Cowley C.R.,  
Castelli F.,\ 2002, A\&A, 387, 595 


\bibitem[Dragon \& Mutschlecner(1980)]{1980ApJ...239.1045D} Dragon 
J.N., Mutschlecner J.P.,\ 1980, ApJ, 239, 1045 

\bibitem[Fuhrmann, Axer, \& Gehren(1993)]{1993A&A...271..451F} Fuhrmann 
K., Axer M., Gehren T.,\ 1993, A\&A, 271, 451 

\bibitem[Gardiner, Kupka, \& Smalley(1999)]{1999A&A...347..876G} Gardiner
R.B., Kupka F., Smalley B.,\ 1999, A\&A, 347, 876 

\bibitem[Gray(1992)]{1992oasp.book.....G} Gray D.F.,\ 1992, The observation
and analysis of stellar photospheres, Cambridge 
Astrophysics Series, Cambridge, Cambridge University Press, 2nd ed.
 
\bibitem[Griffin \& Lynas-Gray(1999)]{1999AJ....117.2998G} Griffin 
R.E.M., Lynas-Gray A.E.,\ 1999, AJ, 117, 2998 


\bibitem[Hinkle, Wallace, Valenti, \& Harmer(2000)]{2000vnia.book.....H} 
Hinkle K., Wallace L., Valenti J., Harmer D.,\ 2000, in
Hinkle, K., Wallace, L., Valenti, J., Harmer D., eds, 
Visible and near 
infrared atlas of the Arcturus spectrum 3727-9300 \AA, San Francisco, ASP 

\bibitem[]{} Hubeny I., 1988, Comp. Phys. Comm., 52, 103

\bibitem[]{} Hubeny I.,  Lanz T., 2000, Synspec -- A User's Guide

\bibitem[Hubeny, Hummer, \& Lanz(1994)]{1994A&A...282..151H} Hubeny I., 
Hummer D.G., Lanz T.,\ 1994, A\&A, 282, 151 

\bibitem[Kupka et al.(1999)]{1999A&AS..138..119K} Kupka F., Piskunov N., 
Ryabchikova T.A., Stempels H.C., Weiss W.W.,\ 1999, A\&AS, 138, 119 


\bibitem[Kurucz(1993)]{1993KurCD..13.....K} Kurucz R.,\ 1993, ATLAS9 
Stellar Atmosphere Programs and 2 km/s grid.~Kurucz CD-ROM No.~13.~ 
Cambridge, Mass.: Smithsonian Astrophysical Observatory  


\bibitem[Kurucz, Furenlid, \& Brault(1984)]{1984sfat.book.....K} Kurucz 
R.L., Furenlid I., Brault J.,  Testerman L., \ 1984, National Solar Observatory Atlas, 
Sunspot, New Mexico: National Solar Observatory


	
\bibitem[Morgan, Keenan, \& Kellman(1943)]{1943QB881.M6.......} Morgan 
W.W., Keenan P.C.,  Kellman E.,\ 1943, , The University 
of Chicago press, Chicago, Illinois

\bibitem[Prugniel \& Soubiran(2001)]{2001A&A...369.1048P} Prugniel P.,
Soubiran C.,\ 2001, A\&A, 369, 1048 


\bibitem[Snider et al.(2001)]{2001ApJ...562..528S} Snider S., Allende 
Prieto C., von Hippel T., Beers T.C., Sneden C., Qu Y.,  Rossi S.,\ 
2001, ApJ, 562, 528 

\bibitem[Th{\' e}venin \& Idiart(1999)]{1999ApJ...521..753T} Th{\' e}venin 
F., Idiart T.P.,\ 1999, ApJ, 521, 753 


\end{thebibliography}
\end{document}